\documentclass[twocolumn,showpacs,preprintnumbers,floats,aps,pra,10pt]{revtex4-1}
\usepackage{graphicx}
\usepackage{dcolumn}
\usepackage{amsmath}
\usepackage{amsfonts}
\usepackage{amssymb}
\usepackage{dcolumn}
\usepackage{color}
\usepackage{float}
\usepackage{subfig}
\usepackage{tabularx}
\usepackage{soul}
\usepackage{ulem}

\newcolumntype{L}[1]{>{\raggedright\arraybackslash}p{#1}}
\newcolumntype{C}[1]{>{\centering\arraybackslash}p{#1}}
\newcolumntype{R}[1]{>{\raggedleft\arraybackslash}p{#1}}

\begin{document}
\title{Constraining a causal dissipative cosmological model}

\author{Norman Cruz}
\altaffiliation{norman.cruz@usach.cl}
\affiliation{Departamento de F\'isica, Universidad de Santiago de Chile, \\
Avenida Ecuador 3493, Santiago, Chile.}

\author{A. Hern\'andez-Almada}
\altaffiliation{ahalmada@uaq.mx}
\affiliation{Facultad de Ingenier\'ia, Universidad Aut\'onoma de Quer\'etaro, \\
Cerro de las Campanas s/n, Santiago de Quer\'etaro, Mexico.}

\author{Octavio Cornejo-P\'erez}
\altaffiliation{octavio.cornejo@uaq.mx}
\affiliation{Facultad de Ingenier\'ia,
Universidad Aut\'onoma de Quer\'etaro, \\
Cerro de las Campanas s/n, Santiago de Quer\'etaro, Mexico.}

\date{\today}

\begin{abstract}
\textbf{{Abstract:}} In this paper a cosmological solution of polynomial type $H \approx ( t + const.)^{-1}$ for the causal thermodynamical approach of Isarel-Stewart, found in \cite{MCruz:2017, Cruz2017}, is constrained using the joint of the latest measurements of the Hubble parameter (OHD) and Type Ia Supernovae (SNIa). Since the expansion described by this solution does not present a transition from a decelerated phase to an accelerated one, both phases can be well modeled  connecting  both  phases  by requiring the continuity of the Hubble parameter at $z=z_{t}$, the  accelerated-decelerated transition redshift. Our best fit constrains the main free parameters of the model to be 
$A_1= 1.58^{+0.08}_{-0.07}$ ($A_2=0.84^{+0.02}_{-0.02}$) 
for the accelerated (decelerated) phase. For both phases we obtain $q=-0.37^{+0.03}_{-0.03}$ ($0.19^{+0.03}_{-0.03}$) and $\omega_{eff} = -0.58^{+0.02}_{-0.02}$ ($-0.21^{+0.02}_{-0.02}$) for the deceleration parameter and the effective equation of state, respectively. 
Comparing our model and  LCDM statistically  through  the  Akaike  information  criterion and  the  Bayesian  information  criterion we obtain that the LCDM model is preferred  by  the OHD+SNIa data. Finally, it is shown that the constrained parameters values satisfy the criterion for a consistent fluid description of a dissipative dark matter component, but with a high value  of  the  speed of  sound  within  the  fluid, which is a drawback for a consistent description of the structure formation. We briefly discuss the possibilities to overcome this problem 
with a non-linear generalization of the causal linear thermodynamics of bulk viscosity and also with the inclusion of some form of dark energy.
\vspace{0.5cm}
\end{abstract}
 \maketitle

\section{Introduction}

The Universe is currently in an accelerated expansion epoch that has been observed through the type Ia supernovae (SNIa) \cite{Riess:1998,Perlmutter:1999}, and the large-scale structure (LSS) \cite{Abbott:2017wau}. Typically, this phenomenon is associated to a component known as dark energy (DE), and together with the one named dark matter (DM), it constitutes the dark sector that corresponds to about $96\%$ of the Universe \cite{Planck:2018}. The simplest cosmological model to explain this dark sector and also compatible with the observational data is the so-called $\Lambda$-Cold Dark Matter ($\Lambda$CDM). This model proposes a cosmological constant as the responsible of the accelerated expansion of the Universe, and a non-relativistic entity without pressure as the dark matter. 
However, one of the open problems in the investigation of the dark sector is its division into DM and DE, which has been proven to be merely conventional since exists a degeneracy between both components, resulting from the fact that gravity only measures the total energy tensor~\cite{Kunz,Sandvik}. So, in the lack of a well confirmed detection (nongravitational) of the DM only the overall properties of the dark sector can be inferred from cosmological data, at the background and
perturbative levels. These results have driven the research to explore alternative models which consider a single fluid that behaves as DM, but also presents the effects of an effective negative pressure at some stage of the cosmic evolution. They are called Unified DM models (UDM) and examples of them are: (Generalized) Chaplygin fluids \cite{Chaplygin,Kamenshchik:2001,Bilic:2001,Fabris:2001}, logotropic dark fluid \cite{Chavanis:2016}, and more recently generalized perfect fluid models \cite{Hova2017, Almada:2018}. Apart of them, exists the possibility of explain the accelerated expansion of the Universe at late times as an effect of the effective negative pressure, due to bulk viscosity in the cosmic fluids, and was first considered in \cite{Padmanabhan:1987, Gron:1990}. Several models regarding this approach have been studied and constrained using cosmological data \cite{Xin-He:2009, XuDou:2011, Velten:2011, Calogero:2013, Normann:2016, Cruz_2018}.

A consistent description of the relativistic thermodynamics of non perfect fluids is the causal description framework given by the Israel-Stewart (IS) theory \cite{Israel1979}.  Due to the high degree of nonlinearity of the differential equations involved, only some exact solution has been found for a simple Ansatz of the bulk viscosity coefficient $\xi$, as a function of the energy density $\rho$ of the fluid with dissipation. For the election $\xi =\xi _{0}\rho ^{1/2}$, a cosmological solution of polynomial type $H \approx ( t + const.)^{-1}$ for the Hubble rate was found as an Ansatz in \cite{MCruz:2017, Cruz2017}, which can describe accelerated, decelerated or even a phantom type cosmic expansion.

This solution can also be obtained in a systematic way by applying the factorization method to the dynamics equation for the Hubble rate.  The factorization of second-order linear ordinary differential equations (ODEs), is a well established method to get solutions in an algebraic manner. It goes back to some works of Dirac to solve the spectral problem for the quantum oscillator \cite{dirac1}, and a further development due to Schrodinger's works on the factorization of the Sturm-Liouville equation \cite{schro1,schro2}. However, in recent times, the factorization technique has been developed and applied to find exact solutions of nonlinear second-order ODEs \cite{berkovich1,cornejo1,rosu1,wang1,rosu3,tiwari:2015,hazra:2012}. The basic concept follows the same pattern already used in linear equations, and it works efficiently for ODEs with polynomial nonlinearities. The method is well adapted to the Hubble rate ODE which raises for instance in viscous cosmological models \cite{Belinchon:2017,cornejo2:2013}.

The main aim of this work is to constrain this solution using the latest measurements of the Hubble parameter (OHD) and Type Ia Supernovae (SNIa), reported in \cite{Magana:2018} and \cite{Scolnic:2017caz}, respectively. Despite the fact that the expansion described by this solution does not present a transition from a decelerated phase to an accelerated one, which is an ultimate feature supported by the observational data, both phases can be well modeled by separated using the analytical solution obtained, as we will show in our results.  

In the case of the non causal Eckart's approach, $\xi _{0}$ can be estimated, for example, directly from the
observational data~ \cite{Avelino2013}. Nevertheless, in the case of our solution, the observational constraints lead to allowed regions for $\xi_{0}$ and the parameter  $\epsilon$, which is related to the non adiabatic contribution to the speed of sound in the viscous fluid, as it will be discussed in Section II. 
Since the above mentioned parameters are involved in a constraint which is a necessary condition for maintaining the thermal equilibrium, we will discuss our results considering such constraint.

This paper is organized as follows: in section \ref{dos}, we describe
briefly the causal Israel-Stewart theory, showing the general differential equation to be solved. In section
\ref{sec:Solution}, we solve this differential equation by using the factorization technique. In section IV, we present the constraints for our model using the observational data coming from the direct measurements of the Hubble parameter and SNIa. Finally, in section V, we discuss our results.

\section{Israel-Stewart-Hiscock formalism} \label{dos}

In what follows we shall present briefly the Israel-Stewart-Hiscock
formalism to describe the thermodynamic properties and evolution of
a Universe filled with only one fluid as the main component, which
experiments dissipative process during its cosmic evolution. We
assume that this fluid obeys a barotropic EoS, $p=\omega \rho $,
where $p$ is the barotropic pressure and $0\leq \omega <1$. For a
flat FLRW Universe, the equation of constraint is
\begin{eqnarray}
3H^{2}=\rho. \label{eq:eq0}
\end{eqnarray}
In the ISH framework, the transport equation for the viscous pressure
$\Pi $ is given by \cite{Israel1979}
\begin{equation}
\tau\dot{\Pi}+ \left(1+\frac{1}{2}\tau\Delta\right)\Pi=-3\xi(\rho) , \label{eqforPi}
\end{equation}
where "dot" accounts for the derivative with respect to the cosmic
time. $\tau$ is the relaxation time, $\xi (\rho)$ is the bulk
viscosity coefficient, for which we assume the dependence upon the energy density $\rho$, $H$ is the Hubble parameter and $\Delta$ is defined by 
\begin{equation}
\Delta = 3H+\frac{\dot{\tau}}{\tau }-\frac{\dot{\xi}}{\xi}-\frac{\dot{T}}{T}, \label{Delta}
\end{equation}
where $T$ is the barotropic temperature, which takes the form
$T=\beta \rho ^{\omega /\left(\omega +1\right)}$ that is the Gibbs
integrability condition when $p=\omega \rho $ and $\beta$ is a
positive parameter. We also have that~\cite{Maartens1996}
\begin{equation}
\frac{\xi}{\left(\rho +p\right)\tau} =
c_{b}^{2},\label{relaxationtime}
\end{equation}
where $c_{b}$ is the speed of bulk viscous perturbations (non-adiabatic contribution to the speed of sound in a dissipative
fluid without heat flux or shear viscosity), $c_{b}^{2}=\epsilon
\left(1-\omega \right)$ and $0<\epsilon \leq 1$, in order to ensure causality, with a dissipative speed of sound lower or equal to the speed of light. We shall also assume a power law dependence for $\xi$ in terms of the energy density of the main fluid, i.e., $\xi =\xi _{0}\rho ^{s}$ where $s$ is an arbitrary
parameter and $\xi _{0}$ a positive constant, in order to satisfy the second law of thermodynamics~\cite{Weinberg1971}. This particular election of $\xi(\rho)$ is rather arbitrary, but allows to obtain a differential equation for the Hubble parameter that can be integrated for some particular values of $s$, obtaining well known analytic
solutions. As we will discuss below, the case $s=1/2$ leads to the most simple form of the differential equation involved.

Using the barotropic EoS in Eq.(\ref{relaxationtime}), we obtain the following expression for the relaxation time 
\begin{eqnarray}
\tau =\frac{\xi_0}{\epsilon (1-\omega ^{2}) }\rho ^{s-1},  \label{eq:eq3}
\end{eqnarray}
and according to Eq. (\ref{Delta})

\begin{eqnarray}
\Delta =\frac{3H}{\delta ( \omega ) }\left( \delta (
\omega ) -\frac{\dot{H}}{H^2}\right) , \label{eq:eq4}
\end{eqnarray}
where we have defined the $\delta ( \omega )$ parameter by

\begin{eqnarray}
\delta \left( \omega \right) \equiv \frac{3}{4}\left( \frac{1+\omega
}{1/2+\omega } \right). \label{eq:eq5}
\end{eqnarray}
So, for $0\leq \omega <1$, $\delta \left( \omega \right) >0$. Using Eqs. (\ref{eq:eq0}) and (\ref{eq:eq3}) we can write

\begin{eqnarray}
\tau H=\frac{3^{s-1}\xi _{0}}{\epsilon (1-\omega ^{2})}H^{2\left( s-1/2\right)
}. \label{eq:eq7}
\end{eqnarray}
For the particular case $s=1/2$ we obtain that 
\begin{eqnarray}
\tau H=\frac{\xi _{0}}{\sqrt{3}\epsilon (1-\omega ^{2})}
. \label{eq:eq8}
\end{eqnarray}
In this case, the necessary condition for keeping the fluid description of the dissipative dark matter component is given by $\tau H < 1$, which leads to the upper limit for  $\xi _{0}$ 
\begin{eqnarray}
\xi _{0} < \sqrt{3}\epsilon (1-\omega ^{2})
. \label{eq:eq9}
\end{eqnarray}
We will discuss later this condition when a cold dark matter fluid with dissipation, as the main component of a late time Universe, be constrained by the observational data.

The differential equation for the Hubble parameter can be constructed by using the conservation equation

\begin{eqnarray}
\dot{\rho}+3H\left[ \left( 1+\omega \right) \rho +\Pi \right] =0,
\label{eq:eq18}
\end{eqnarray}
the Eqs. (\ref{eq:eq0}) and (\ref{eqforPi}), and the relation $\xi \left( \rho \right) =\xi
_{0}\rho ^{s}$. So, we can obtain the following
differential equation
\begin{widetext}
\begin{eqnarray}
\left[ \frac{2}{3\left( 1-\omega ^{2}\right) }\left( \frac{3\left(
1+\omega \right)\dot{H}}{H^{2}}+\frac{\ddot{H}}{H^{3}}\right)
-3\right] H^{2\left( s-1/2\right) }+ \frac{1}{3^{s}\xi _{0}}\left[
1+\frac{3^{s-1}\xi _{0}\Delta H^{2\left( s-1\right)}}{2\left(
1-\omega ^{2}\right) } \right] \left[ 3\left( 1+\omega \right)
+\frac{2\dot{H}}{H^{2}}\right] = 0. \label{eq:eq19}
\end{eqnarray}
\end{widetext}
For $s=1/2$, the following Ansatz
\begin{eqnarray}
H\left( t\right) =A\left( t_{s}-t\right) ^{-1}, \label{Ansatz}
\end{eqnarray}
is a solution of Eq.(\ref{eq:eq19}) with a big rip singularity \cite{Cruz2017}, and the Ansatz
\begin{eqnarray}
H\left( t\right) =A\left( t -t_{s}\right) ^{-1}, \label{Ansatz1}
\end{eqnarray}
is also a solution which can describe cosmic evolutions with accelerated, linear and decelerated expansion 
\cite{MCruz:2017}. In the next section, we will show that by using the factorization method this Ansatz can be obtained as a particular solution of the differential equation (\ref{eq:eq19}), which gives a deeper understanding of its particularity and its dependence on the initial conditions.


\section{Solving the differential equation for the Hubble rate} \label{sec:Solution}

The nonlinear differential equation for the Hubble function (\ref{eq:eq19})
can be rewritten for $s=1/2$ as follows
\begin{equation}\label{facto1}
 \ddot{H}  + \frac{\alpha_1}{H} {\Dot H}^2  + \alpha_2 H {\Dot H} + \alpha_3 H^3=0,
\end{equation}
where
\begin{eqnarray}
\alpha_1 & = & -\frac{3}{2\delta},\\
\alpha_2 & = & \frac{3}{2} + 3(1+\omega)-\frac{9}{4\delta}(1+\omega)+\frac{\sqrt{3}\epsilon(1-\omega^2)}{\xi_0},\\
\alpha_3 & = & \frac{9}{4}(1+\omega) + \frac{9}{2}\epsilon (1-\omega^2)\left[ \frac{1+\omega}{\sqrt{3}\xi_0} -1 \right],
\end{eqnarray}
are constant coefficients. 

Let us consider the following factorization scheme \cite{cornejo1,rosu1,wang1} to obtain an exact particular solution of the Eq. (\ref{facto1}).
The nonlinear second order differential equation
\begin{equation}\label{facto1a}
\ddot{H} + f(H)\dot{H}^2 + g(H)\dot{H} + j(H)=0,
\end{equation}
where $\dot{H}=\frac{dH}{dt}=D_t H$, can be factorized in the form
\begin{equation}\label{facto2}
 [D_t - \phi_1(H)\dot{H} - \phi_2(H)][D_t - \phi_3(H)]H=0.
\end{equation}
where $\phi_i(H)$ ($i=1,2,3$) are factoring functions to be found. Expanding Eq. (\ref{facto2}), one is able to group terms as follows \cite{wang1}
\begin{equation}\label{facto2a}
\ddot{H} -\phi_{1}\dot{H}^2 + \left( \phi_{1}\phi_{3}H - \phi_{2}-\phi_{3}-\frac{d\phi_{3}}{dH}H\right) \dot{H}+\phi_{2}\phi_{3}H =0.
\end{equation}
Then, by comparing Eq. (\ref{facto1a}) with Eq. (\ref{facto2a}), we get the following conditions%
\begin{align}
f\left(  H\right)  &  = -\phi_{1},\label{facto3}\\
g\left( H \right)  &  = \phi_{1}\phi_{3}H - \phi_{2}-\phi_{3}-\frac{d\phi_{3}}{dH}H,\label{facto4}\\
j\left(H \right)  &  =\phi_{2}\phi_{3}H.\label{facto5}
\end{align}
Any factorization like (\ref{facto2}) of an scalar ODE in the form given in (\ref{facto1a}) allows to find a compatible first order ODE \cite{berkovich1}
\begin{equation}
[D_t - \phi_3(H)]H= D_tH - \phi_3(H)H=0,\label{facto5a}
\end{equation}
whose solution provides a particular solution of Eq. (\ref{facto1a}).

We apply now the previous scheme to Eq. (\ref{facto1}). The factoring function $\phi_1=-\frac{\alpha_{1}}{H}$ since $f(H)$ is explicitly given in Eq. (\ref{facto1}). Also, according to Eq. (\ref{facto5}) the two unknown functions $\phi_2$ and $\phi_3$ are easily obtained by merely factoring the polynomial expression $j(H)=\alpha_3 H^3$ given as well in Eq. (\ref{facto1}).
Then, the functions
\begin{equation}
\phi_{2}=a_{1}^{-1}H,\quad\text{and}\quad
\phi_{3}=a_{1}\alpha_3 H,\label{facto6}
\end{equation}
where $a_{1}(\neq0)$ is an arbitrary constant, are proposed.

The explicit value of $a_{1}$ is obtained by substituting $ g(H)= \alpha_2 H$ and the $\phi_i$ functions into Eq. (\ref{facto4}). Then, we get the constraint equation 
\begin{equation}
 \alpha_2 H = -(a_1\alpha_1 \alpha_3 + a_1^{-1} +2a_1\alpha_3 )H, \label{facto6a} 
\end{equation}
and equating both sides of the equation provides
\begin{equation}
a_1 = \frac{-\alpha_2 \pm \sqrt{\alpha_2^2 - 4\alpha_3\left(2+\alpha_1\right)}}{2\alpha_3 \left( 2 + \alpha_1 \right) }. \label{facto6b}
\end{equation}

Therefore, the Eq. (\ref{facto1}) admits the factorization 
\begin{equation}
\left[ D_t + \frac{\alpha_1}{H}\dot{H} -a_1^{-1}H\right]\left[D_t -a_1 \alpha_3 H\right]H = 0,\label{facto8}
\end{equation}
with the compatible first order ODE
\begin{equation}
\dot{H} -a_1 \alpha_3 H^2 = 0,\label{facto9}
\end{equation}
whose solution is also a particular solution of the Eq. (\ref{facto1}) factorized in the form (\ref{facto8}).

The integration of this equation generates one arbitrary integration constant, which can be written in explicit terms of an initial condition. If we consider the initial condition $H(t_0)=H_0$, where $H_0$ is the Hubble constant, then we get the following particular solution of Eq. (\ref{eq:eq19}) with $s=1/2$,
\begin{equation}
H(t)= \frac{A_{\pm}}{t-(t_0-\frac{A_{\pm}}{H_0})}, \label{facto10}
\end{equation}
where $A_{\pm}= -\frac{1}{\alpha_3 a_1}$, 
or equivalently
\begin{equation}
A_{\pm}= \frac{2\sqrt{3}\epsilon(\omega^2-1)-6\xi_0\pm 2\sqrt{3\epsilon}\sqrt{\epsilon(\omega^2 -1)^2 +6\xi_0^2(1-\omega)}}{3(\omega+1)\left( -3\xi_0+2\epsilon (\omega -1)\left[ \sqrt{3}(1+\omega) - 3\xi_0 \right] \right)},
\label{facto25}
\end{equation}
with the restriction equation $\xi_0 \neq \frac{2\sqrt{3}\epsilon\left(\omega^2 -1 \right)}{3+6\epsilon\left(\omega-1\right)}$, which avoids $A_{\pm}$ to be an indeterminate function.
The above particular solution (\ref{facto10}) can also be written in the form
\begin{equation}
H(t)= \frac{A_{\pm}}{t-t_s}, \label{facto13}
\end{equation}
where $t_s=t_0-\frac{1}{H_0(1+q_0)}$, and $q_0$ is the initial value of the deceleration parameter, but since
\begin{equation}\label{eq:q}
1+q = -\frac{\dot{H}}{H^2} = \frac{1}{A_{\pm}} \,,
\end{equation}
this means that this solution represents an expansion with a constant deceleration parameter.  Once a $q_0$ is given, a value is obtained for $A_{\pm}$ and a family of possible values for the parameters $\epsilon$, $\omega$ and $\xi_0$ can be evaluated from Eq. (\ref{facto25}). Or, once the value of $A_{\pm}$ is given, or constrained from the data, as it will be done in the next section,  $q_0$ and the other ranges of the parameters can be evaluated. 

The solution (\ref{facto13}) can also be written in terms of the redshift variable. For the scale factor one obtains
\begin{equation}
    \frac{a}{a_0}= \left( \frac{t-t_s}{t_0-t_s} \right)^{A_{\pm}} = \frac{1}{1+z}. \label{facto14}
\end{equation}
Therefore, 
\begin{equation}
    H(z)= H_0 (1+z)^{1/A_{\pm}} \,, \label{facto15}
\end{equation}
where $H_0 = 100\,h\,\, km s^{-1} Mp c^{-1}$, and $h$ denotes the dimensionless Hubble constant.  Notice that this form of the Hubble parameter is defined for both phases of the Universe, the accelerated and decelerated one. However, we can connect both phases by requiring the continuity of the Hubble parameter function at $z=z_t$, where $z_t$ is the accelerated-decelerated transition redshift. Then, we obtain

\begin{equation} \label{eq:Hz}
H(z)= \left\{ \begin{array}{lc}
             H_0(1+z)^{1/\hat{A}_1}\,,    & z \leq z_t\,, \\
             \\ 
             H_0 (1+z_t)^{1/\hat{A}_1 - 1/\hat{A}_2}(1+z)^{1/\hat{A}_2}\,,   & z > z_t\,. \\
             \end{array}
      \right.
\end{equation}
In the above expression, $\hat{A}_1$ and $\hat{A}_2$ are the free parameters corresponding to the accelerated and decelerated phases respectively.

\section{Cosmological constraints}

In this section we describe the observational data used and build the $\chi^2$-function to perform the confidence regions of the free model parameters. 
We employ a Chain Markov Monte Carlo 
analysis based on emcee module \cite{Emcee:2013} by setting $5000$ chains with $500$ steps. 
The nburn is stopped up to obtain a value of $1.1$ on each free parameter in the Gelman-Rubin criteria \cite{Gelman:1992}.
Table \ref{tab:priors} presents the priors considered for each parameter. We also set the redshift of the accelerated-decelerated transition as $z_t=0.64$ \cite{Moresco:2016mzx} in the Eq. (\ref{eq:Hz}). Then, in order to constrain the model parameters we use the Hubble parameter measurements and supernovae data, and the combined data.
\begin{table}
\caption{Priors considered for the model parameters.}
\centering
\begin{tabular}{| C{2cm}  C{3cm} |}
\hline
Parameter & Prior       \\
\hline
$\hat{A}_1$   & Flat in $[1,5]$ \\ [0.7ex]
$\hat{A}_2$   & Flat in $[0,1]$ \\ [0.7ex]
$h$           & Gaus$(0.7324,0.0174)$   \\ [0.7ex]
\hline
\end{tabular}
\label{tab:priors}
\end{table}

\subsection{Hubble Observational Data}
The direct way to observe the expansion rate of the Universe is through measurements of the Hubble parameter (OHD) as a function of the redshift,  $H(z)$. The latest OHD obtained by using the differential age (DA) method \cite{Jimenez:2001gg}, are compiled in \cite{Magana:2018} and consist of $51$ Hubble parameter points covering the redshift range $[0,1.97]$. We constrain the free model parameters by minimizing the chi-square function 
\begin{equation}\label{eq:chi2_ohd}
\chi^2_{OHD} = \sum_{i} \left( \frac{H_{th}(z_i) - H_{obs}}{\sigma_{obs}^i} \right)^2,
\end{equation}
where $H_{th}(z_i)$ and $H_{obs}(z_i) \pm \sigma_{obs}^i$ are the theoretical and observational Hubble parameter at the redshift $z_i$, respectively.

\subsection{Type Ia Supernovae}
We use the Pantheon dataset \cite{Scolnic:2017caz} consisting of $1048$ type Ia supernovae (SNIa) located into the range $0.01 < z < 2.3$. The comparison between data and model is obtained with the expression
\begin{equation}\label{eq:chi2_sn}
\chi^2_{SNIa} = (m_{th}-m_{obs}) \cdot {\rm Cov}^{-1} \cdot (m_{th}-m_{obs})^{T}
\end{equation}
where $m_{obs}$ is the observational bolometric apparent magnitude and ${\rm Cov}^{-1}$ is the inverse of the covariance matrix. $m_{th}$ is the theoretical estimation and is computed by
\begin{equation}
m_{th}(z) = \mathcal{M} + 5\, \log_{10}\left[ d_L(z)/10\,pc \right]\,.
\end{equation}
Here, $\mathcal{M}$ is a nuisance parameter and $d_L(z)$ is the dimensionless luminosity distance given by
\begin{equation}
d_L(z) =  (1+z)\,c \int_0^z \frac{dz'}{H(z')}
\end{equation}
where $c$ is the speed of light.

\subsection{Joint analysis}
We also perform a joint analysis by defining the merit-of-function as
\begin{equation}
\chi^2_{joint} = \chi^2_{OHD} + \chi^2_{SNIa}\,,
\end{equation}
where $\chi^2_{OHD}$ and $\chi^2_{SNIa}$ are given in Eqs. (\ref{eq:chi2_ohd}) and (\ref{eq:chi2_sn}), respectively. 
The best fitting parameters are obtained by setting the acceleration-deceleration transition $z_t=0.64$ \cite{Moresco:2016mzx}. Table \ref{tab:bf_values} presents the summary of the best estimates of the parameters for the dissipative unified dark matter (DUDM) model (see Eq. (\ref{eq:Hz})).

\begin{table*}
\caption{Best fit values of the free parameters of the UDM model.}
\centering
\begin{tabular}{|c c c c c c c c|}
\hline
Data    & $\chi^2$ &    $\hat{A}_1$ & $\hat{A}_2$   & $h$    & $\mathcal{M}$ &  BIC & AIC     \\
\hline 
OHD      & $34.10$  & $1.58^{+0.15}_{-0.12}$ & $0.84^{+0.02}_{-0.02}$ & $0.700^{+0.014}_{-0.014}$ & - & $57.69$  & $40.10$     \\ [0.7ex]
SNIa     & $1029.48$  & $1.62^{+0.11}_{-0.10}$ & $0.71^{+0.16}_{-0.14}$ & $0.732^{+0.017}_{-0.017}$& $5.76^{+0.05}_{-0.05}$ & $1085.12$ & $1035.48$    \\ [0.7ex]
OHD+SNIa & $1064.91$  & $1.58^{+0.08}_{-0.07}$ & $0.84^{+0.02}_{-0.02}$ & $0.700^{+0.010}_{-0.010}$& $5.67^{+0.02}_{-0.02}$ & $1120.93$ & $1070.91$      \\ [0.7ex]
\hline
\end{tabular}
\label{tab:bf_values}
\end{table*}

\begin{figure}
  \centering
  \includegraphics[width=.9\linewidth]{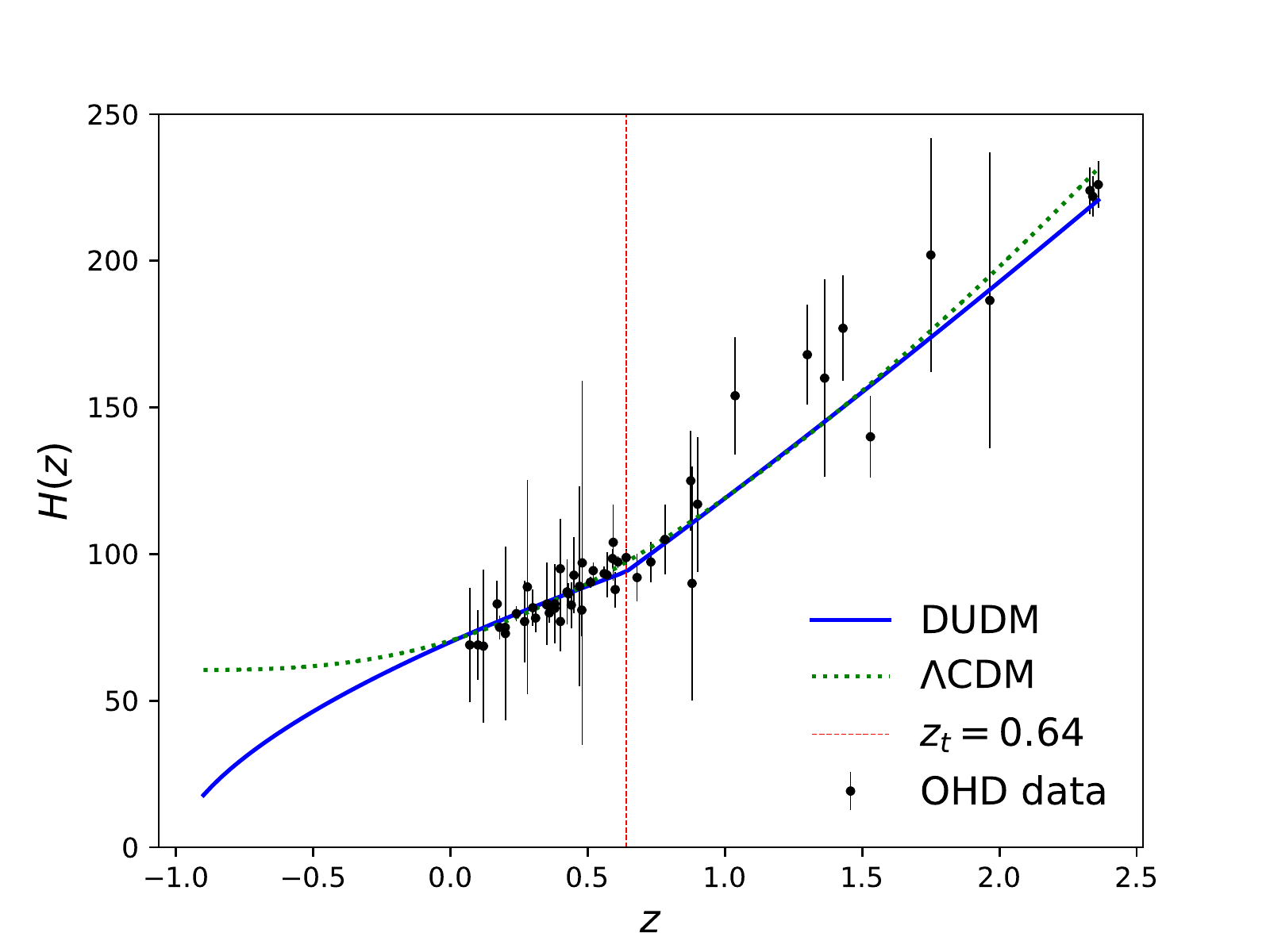}   \\
  \includegraphics[width=.9\linewidth]{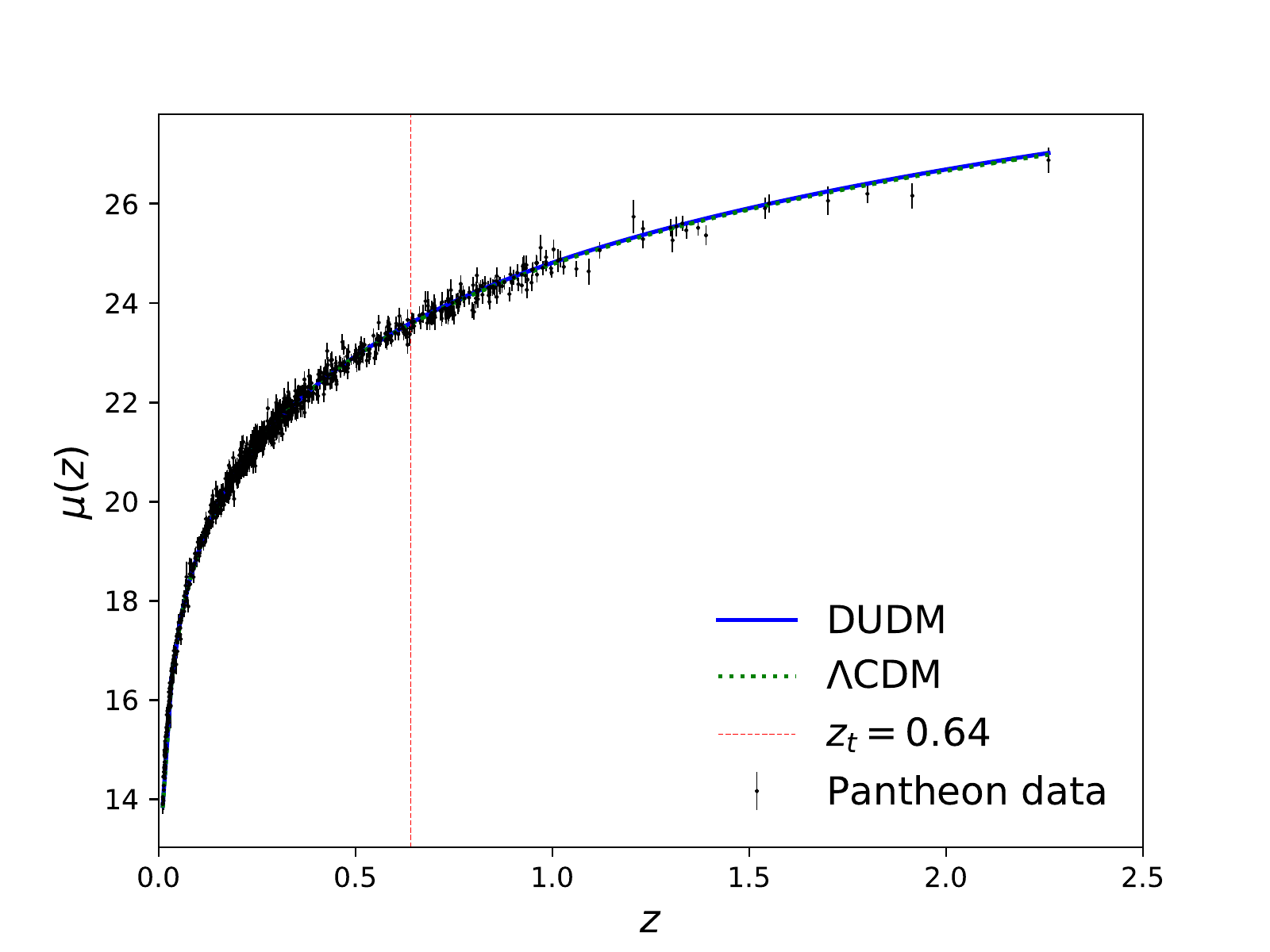}
\caption{Joint best fit of DUDM and $\Lambda$CDM using the best fitting values of joint analysis}
\label{fig:plotHz}
\end{figure}

Figure \ref{fig:plotHz} shows the best fit curves over OHD and SNIa samples at top and bottom panel, respectively, using the joint analysis. We observe an evident behaviour in the Hubble parameter between the DUDM and LCDM at $z<0$ (the future). While LCDM gives an Universe expansion smoothly, the DUDM model has an Universe expansion as big rip. From Eq. (\ref{eq:q}) and the joint analysis values, we estimate the decelerate parameter $q = -0.37$ and $0.19$ for the accelerated and decelerated phases, respectively. Notice that $q$ is constant during each phase.

\begin{figure}
  \centering
  \includegraphics[width=.9\linewidth]{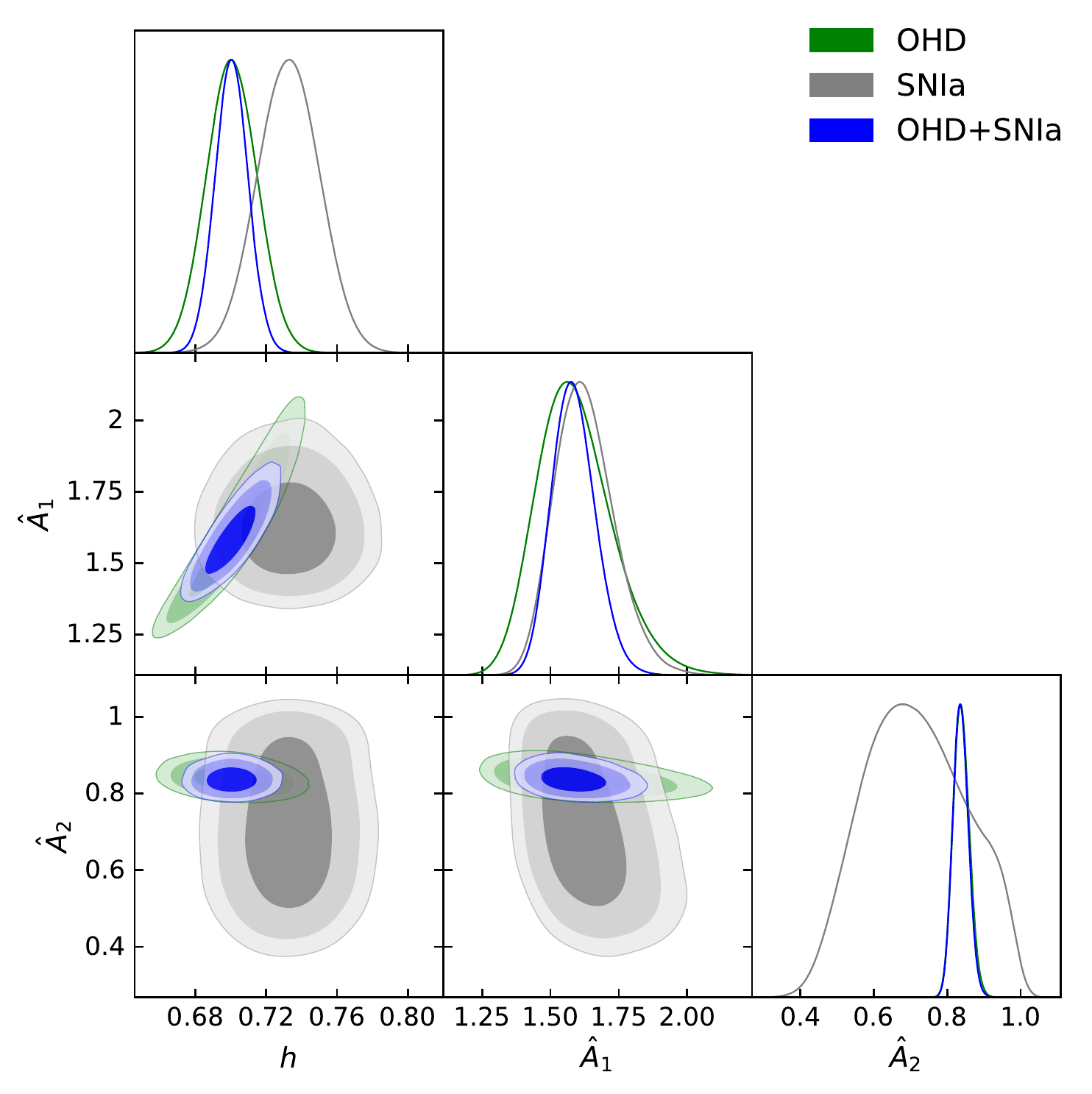}
\caption{2D-contours considering OHD (green), SNIa (gray), and joint analysis (blue)  at $68,95,99.7\,\%$ confidence level.}
\label{fig:Contours}
\end{figure}

Figure \ref{fig:Contours} shows the 2D contours at $68$, $95$ and $99.7\,\%$ ($1,2$ and $3 \sigma$) confidence level (CL) and the 1D posterior distributions of the free model parameters. It shows a good agreement between the best fits within $1\sigma$ CL.

\section{Discussion}

\begin{figure}
  \centering
  \includegraphics[width=.9\linewidth]{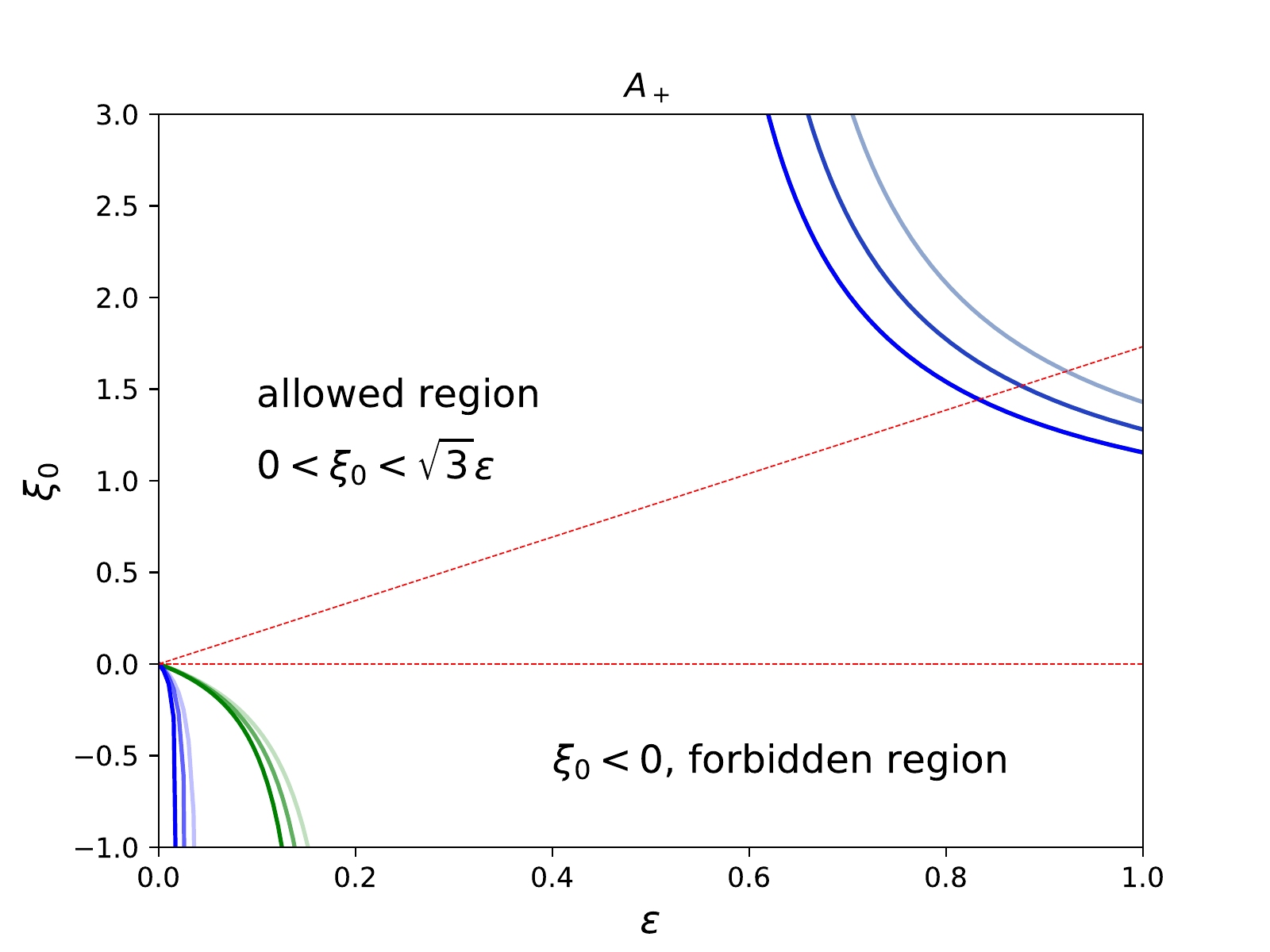}
  \includegraphics[width=.9\linewidth]{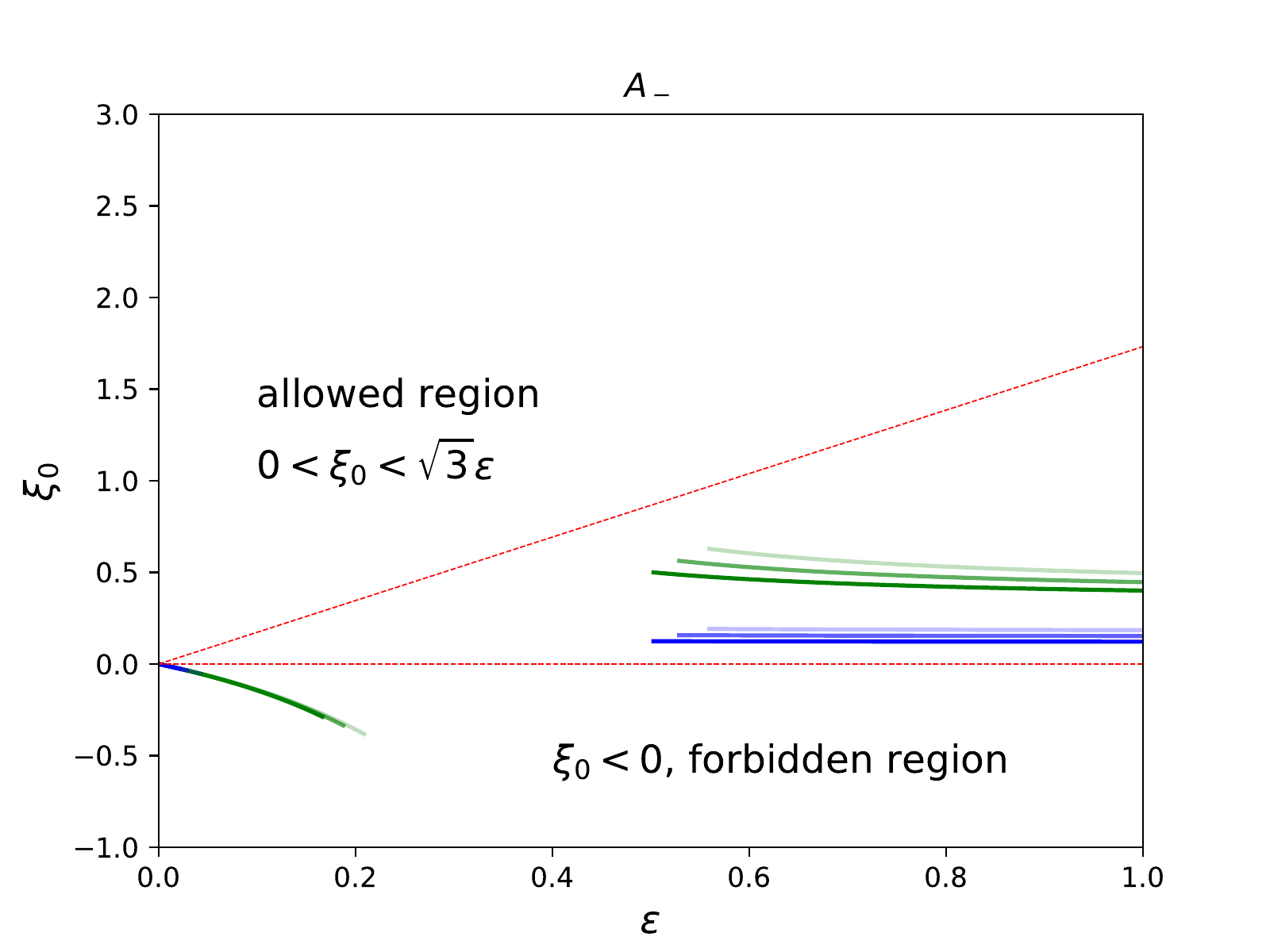}
\caption{Top (bottom) panel displays the behavior of $\xi_0$ as a function of $\epsilon$ considering the positive (negative) sign of the Eq. (\ref{facto25}). The green (blue) color lines correspond to the $\hat{A}_1$ ($\hat{A}_2$) value. In the top panel, the green and blue lines in the region $\xi_0>0$ are superimposed. For both plots and each color, from bottom to top the green (blue) lines refer to $\omega = 0, 0.05, 0.1$, respectively.}
\label{fig:xi0vsEps}
\end{figure}

In the following we will refer the mathematical expressions given in Eq. (\ref{facto25}) as $A_+$ or $A_-$, and the numerical values of each expression could be $\hat{A}_1$ or $\hat{A}_2$.
By using Eq. (\ref{facto25}), which gives $A_\pm$ as a function of the model parameters, we explore the behavior of $\xi_0$ as a function of $\epsilon$. These curves are shown in Figure \ref{fig:xi0vsEps} for several values of $\omega = 0, 0.05,0.1$ (from bottom curve to the top one), and are obtained when we consider the positive (top panel) and negative (bottom panel) sign in Eq. (\ref{facto25}), {\it i.e.}, $A_+$ and $A_-$ respectively. For $A_+$ we find values $\xi_0>0$ in the region $0.5<\epsilon<1$, and $\xi_0<0$ for $0<\epsilon<0.5$. Similarly, when we consider $A_-$, we find positive values of $\xi_0$ in the allowed region $\xi_0<\sqrt{3}\epsilon$ in  $0.5<\epsilon<1$ for both epochs. In contrast, we find values of $\xi_0<0$ within $0<\epsilon<0.5$ for both epochs when any sign is considered.
Then, we discard the phase space of ${\epsilon, \xi_0}$  where $\xi_0<0$ because the second law of thermodynamics would be infringed. It is interesting to note that when we use $A_+$ and $\xi_0>0$, the curves for decelerated/accelerated epochs are not sensitive of $\hat{A}_{1,2}$ values (see Table \ref{tab:bf_values}). 

A further insight of the previous results can be done considering the effective EoS, $\omega_{eff}$, which is defined by  
\begin{equation}\label{eoseff}
\omega_{eff} = -1 - \frac{2}{3}\frac{\dot{H}}{H^2} = -1 + \frac{2}{3}\frac{1}{\hat{A}}.
\end{equation}
where $\hat{A}$ takes the values $\hat{A}_1$ or $\hat{A}_2$ for accelerated or decelerated phase respectively,  For the accelerated phase, we take, $\hat{A}_1= 1.58^{+0.08}_{-0.07}$, corresponding to the obtained value using  OHD+SNIa data. In this case we obtain from Eq. (\ref{eoseff}) that $\omega_{eff}= -0.58^{+0.02}_{-0.02}$, 
which means that the dissipative effects drive a quintessence like behavior.  For the same set of data, $\hat{A}_2= 0.84^{+0.02}_{-0.02}$ in the decelerated phase and $\omega_{eff}= -0.21^{+0.02}_{-0.02}$, therefore in this case even with dissipative effects present in the dark matter fluid, they are not enough to drive acceleration. We find a deviation of $6.2\sigma$ over the region of quintessence ($\omega<-1/3$).

Despite the fact that the solution found does not display a smoothly transition in the deceleration parameter, it allows us to describe both phases separately by using such solution through  the parameters $\hat{A}_1$ and $\hat{A}_2$, derived from cosmological data. 

On the other hand, the condition to keep the fluid description of the dark matter component, which is an essential assumption of the thermodynamical formalism invoked,  given by Eq. (\ref{eq:eq9}) provides the upper limit  $\xi _{0}< \sqrt{3}\epsilon$ for a  pressureless dark matter fluid. By simple inspection of the curves displayed in Figure \ref{fig:xi0vsEps}, it is easy to see that the constraint can be satisfied by both solutions $A_{+}$ and $A_{-}$ and also in the case of the decelerated and accelerated expansions. Nevertheless, this constraint is fulfilled for approximately $\epsilon > 0.82$ for the  $A_{+}$ solution, and for $\epsilon > 0.5$ in the  $A_{-}$ solution. This fact indicates that it is needed a great non adiabatic contribution to the speed of sound within the fluid. It is well known that the structure formation observed implies a very low speed of sound, consistent with a cold dark matter component. Therefore, this issue represents a weakness of the model. Moreover, in the case of the solution with accelerated expansion, the thermal equilibrium of the fluid can not be maintained.  Besides, a positive entropy production and the convexity condition, ${d^2}S/d{t^2}< 0$, are only satisfied by the decelerated solution, as it was shown in \cite{MCruz:2017}.

The solution analyzed in this work takes the simple form given by Eq. (\ref{facto13}), which is too simple and clearly not a general solution of the IS formalism. In fact, only one initial condition is enough to determine the solution, and since it represents a cosmic expansion with a constant deceleration parameter, the other initial condition, $q_0$, necessary to determine the solution of a second order differential equation in the Hubble parameter, plays no role at all. It is hoped that more general solutions could overcome the above spotlighted difficulties. 

Finally, we compare the DUDM and LCDM statistically through the Akaike information criterion (AIC) \cite{AIC:1974, Sugiura:1978}, and the Bayesian information criterion (BIC) \cite{schwarz1978}. The AIC and BIC are defined by ${\rm AIC} = \chi^2 + 2k$ and ${\rm BIC} = \chi^2 + 2k\, \ln(N)$, respectively, where $\chi^2$ is the $\chi^2$ function, $k$ is the number of degree of freedom and $N$ is the data size. The preferred model by data is the one with the minimum value of these quantities. In order to compare the models, we use the full data sample, OHD$+$SNI, and obtain the $\chi^2$ as the sum of the ones obtained in the decelerated and accelerated phases for the DUDM model. Then, we estimate a yield value of $\Delta AIC = AIC^{DUDM}-AIC^{LCDM} = 7.96$ and  $\Delta BIC = BIC^{DUDM}-BIC^{LCDM} =19.96$, which suggest that the LCDM is the model preferred by the OHD$+$SNIa data used. This result is expected since the DUDM model contains a degree of freedom greater than LCDM.

In summary, we analyze an exact solution of a DUDM model using the most recent cosmological data of the Hubble parameter and SNIa, that cover the redshift region $0.01<z<2.3$. Although the exact solution under study was proposed as Ansatz in \cite{MCruz:2017, Cruz2017}, we are able to obtain it in a systematic way by following a factorization procedure \cite{cornejo1,rosu1,wang1}. 
Due to the inability of the model to drive accelerated and decelerated phases with the same value of the main free parameters $A$ as is shown in Eq. (\ref{facto15}), we build the Hubble parameter of the model by connecting both phases as is expressed in Eq. (\ref{eq:Hz}) and being now the free parameters $\hat{A}_+$ and $\hat{A}_-$. Then, we employ an analysis using the combined data, OHD$+$SNIa, and considering the transition redshift  $z_t=0.64$ \cite{Moresco:2016mzx} to constrain their values. According to Eq. (\ref{eq:Hz}) and (\ref{eq:q}), the model presents an acceleration (deceleration) phase when $\hat{A}_1>1$ ($\hat{A}_2<1$). In these epochs, we infer a constant value of $q=-0.37^{+0.03}_{-0.03}$ ($0.19^{+0.03}_{-0.03}$), and an effective EoS $\omega_{eff} = -0.58^{+0.02}_{-0.02}$ ($-0.21^{+0.02}_{-0.02}$). It is interesting to see that $\omega_{eff}$ is in the quintessence region for the accelerated epoch, while the decelerated phase is characterized by a negative effective EoS, even though it is not enough to drive an accelerated expansion of the Universe. We have also found that our solution can well fitted the cosmological data, and the evaluated values of $\xi _{0}$ from the constrained values of $A_{+}$ and $A_{-}$ always satisfy the condition of a fluid description for both phases, required from the thermodynamics formalism. Nevertheless, the high value of the speed of sound within the fluid is an undesirable behavior of the model. 

It is important to point out that our solution is obtained assuming a Universe filled with only one fluid with dissipation, therefore it is clear that it can describe only the late time evolution. An extension of this model to early ages of the Universe requires to introduce radiation and evaluate the behavior of the linear perturbations. In the framework of the Eckart theory, the discussion of the linear perturbations has been realized, for example, in \cite{Barrow:2009}. The found results indicate that viscous dark matter leads to modifications of the large-scale CMB spectrum, weak
lensing and CMB-galaxy cross-correlations, which implies difficulties in order to fit the astronomical data. In the case of a perturbative study in the framework of the causal thermodynamics it was found in \cite{Piattella:2011} that numerical solutions for the gravitational potential seem to disfavour causal theory, whereas the truncated theory leads to results similar to those of the $\Lambda$CDM model for a very small bulk viscous speed.

Let us discuss here what can be a possible way to overcome this difficulty, which is present in this type of models. As we mentioned above, the division into DM and DE is merely conventional due to the degeneracy between both components, resulting from the fact that  gravity only measures the total energy tensor. In the case of DUDM models, the viscous stress provide the negative pressure which allows accelerated phases, but the near equilibrium condition demanded in the thermodynamics approaches of relativistic viscous fluids, implies that the viscous stress must be lower than the equilibrium pressure of the fluid. In general, this condition is not fulfilled and one possibility is to go further and to consider a non-linear generalization of the causal linear thermodynamics of bulk viscosity, where deviations from equilibrium are allowed (see, for example, \cite{CRUZ2017159}). Other possibility is consider a cosmological scenario with dissipative DM and some other DE component. In 
\cite{Cruz_2018}, an introduction of a cosmological constant is considered together with a dissipative DM component. This also allows in some regions to satisfy the near equilibrium condition. Of course, in this scenario UDM models with dissipation are abandoned as consistent models to describe the evolution of the Universe, and, on the other hand, we are assuming the division into DM and DE.

As a conclusion of the above discussion we can say that the solution found within the full causal Israel-Stewart-Hiscock formalism indicates that  accelerated expansion compatible with OHD and SNIa data, can be obtained with only one dissipative DM component, but having a great non adiabatic contribution to the speed of sound within the fluid, which is not compatible with the structure formation. Further investigations are required to solve this drawback including some form of DE, along with the dissipative component.

\bigskip

\section*{Acknowledgments}
The authors acknowledge an anonymous referee for important suggestions in order to improve the presentation of the paper. This work was supported by the Universidad de Santiago de Chile, USACH, through Proyecto DICYT N$^{\circ}$ 041831CM (NC), Vicerrector\'ia de Investigaci\'on, Desarrollo e Innovaci\'on. NC acknowledges the hospitality of the Facultad de Ingenier\'ia, Universidad Aut\'onoma de Quer\'etaro,  M\'exico, where part of this work was done. OCP would like to thank warm hospitality and financial support during a summer research stay at Department of Physics, USACH, and also thanks PRODEP project, M\'exico, for resources and financial support. AHA thanks SNI Conacyt and Instituto Avanzado de Cosmolog\'ia (IAC) collaborations. The authors thankfully acknowledge computer resources, technical advise and support provided by Laboratorio de Matem\'atica Aplicada y C\'omputo de Alto Rendimiento from CINVESTAV-IPN (ABACUS), Project CONACYT-EDOMEX-2011-C01-165873. 

\bibliographystyle{unsrt}
\bibliography{main}

\end{document}